\documentclass[12pt]{iopart}
\usepackage[utf8x]{inputenc}
\usepackage{iopams}
\usepackage{graphicx}

\begin{document}

\title{Position Estimation of a Parametrically Driven Optomechanical System}
\author{A Szorkovszky$^1$, AC Doherty$^2$, GI Harris$^1$ and WP Bowen$^1$}
\address{$^1$ Centre for Engineered Quantum Systems, University of Queensland, Australia}
\address{$^2$ Centre for Engineered Quantum Systems, University of Sydney, Australia}
\ead{alexs@physics.uq.edu.au}

\begin{abstract}
We study the position estimation of a mechanical oscillator undergoing both detuned parametric amplification and continuous quantum measurement. This model, which can be utilised to produce squeezed states, is applied to a general optoelectromechanical system. Using a stochastic master equation formalism, we derive general formulae for the reduction in position uncertainty of one quadrature of motion. The filter for extracting the optimal position estimate from the measurement record is derived. We also find that since this scheme does not work far into the back-action dominated regime, implementing resolved-sideband cooling improves the squeezing only marginally.
\end{abstract}

\pacs{42.50.Lc,03.65.Ta,71.36.+c}

\section{Introduction}

When continuously monitoring the position of a mechanical oscillator at finite temperature, three regimes inevitably emerge\cite{doherty}. In the regime in which even the oscillator's Brownian motion cannot be resolved, termed here the \emph{bad measurement} regime, the uncertainty of the position estimate is dominated by thermal noise. In the \emph{classical measurement} regime, the thermal Brownian motion can be resolved but the zero-point motion cannot, and the uncertainty is then dominated by shot-noise. Finally, a measurement strong enough to resolve the zero-point motion results in the \emph{back-action dominated} regime. The border between the second and third regimes, when considering force sensitivity, is usually termed the ``standard quantum limit''(SQL)\cite{caves}. It is here that the uncertainty of the position estimate begins to saturate at the ground state variance, provided the measurement is efficient and suitably filtered\cite{doherty}.

Optomechanical experiments close to and beyond the SQL are presently becoming feasible\cite{anets,hertzberg}, enabling feedback\cite{doherty} or sideband cooling\cite{marquardt} to near the ground state. If a spectroscopic or back-action evading measurement scheme is implemented, this regime also allows one to perform quantum tomography\cite{vanner} or to produce squeezed states\cite{clerk}. Much attention has been paid to such methods which use, at most, modulated measurement signals and linear forces.

Once near the ground state, nonlinear classical forces present a natural alternative way to prepare quantum states of an oscillator. For example, parametric amplification can be used to produce quantum squeezing of one quadrature of motion without requiring back-action evasion\cite{blencowe1}. However, when using ground-state resolving measurement in addition to these nonlinear external forces, one can expect the oscillator dynamics to be dominated by back-action, thus limiting the effect of the external control. As a result, and as we will discuss in detail, a compromise must be made between strong measurement (better cooling) and weak measurement (better control).

Using the model of an optoelectromechanical system, which exhibits the required quantum measurement and classical control\cite{lee,sridaran}, we examine the effect of parametric amplification on quantum measurement (and vice versa) in various scenarios. Parametric amplification has already been demonstrated classically in MEMS and NEMS oscillators\cite{karabalin,rugar,unterreithmeier}, and in the quantum regime for trapped atoms\cite{meekhof} and has long been considered as an alternative to back-action evasion\cite{rugar}. This is due to its nearly limitless ability to ``pre-amplify'' in-phase position fluctuations so that strong measurement is unnecessary for ground-state resolution.

A parametric drive will also squeeze the unconditional thermal noise in the out-of-phase quadrature. In the stable below threshold regime, this squeezing does not exceed a factor of two. This is also true if the drive is detuned from resonance\cite{carmichael,batista}. However, detuning correlates the in-phase and out-of-phase quadratures and therefore becomes more interesting in the context of position estimation. Measurement of the well-resolved amplified quadrature in this case improves the estimation of the squeezed quadrature. This idea has been outlined for the first time recently\cite{prl}, where it was shown that using this approach the potential squeezing is limited only by the oscillator Q. Critically, this paper focused on fixed sub-optimal quadratures and an analytic solution for the squeezing was only derived for the classical measurement regime. Furthermore, experimental implementation of this idea would require the filter --- also absent from the previous work --- which extracts the optimal position estimate from the measurement record. Here, we provide a complete analysis of the system, with additional focus on phenomena related to optomechanics.

In this paper, we provide general analytic solutions for the conditional variance of a parametrically driven oscillator undergoing continuous measurement. We also determine the exact form of the filter which produces an optimal position estimate, where the uncertainty of this estimate is given by the conditional variance. Since the quantum squeezing is ultimately limited by the temperature of the oscillator, it is natural to ask whether combining this scheme with resolved sideband cooling allows an increased amount of squeezing. To that end we introduce a master equation that includes resolved sideband cooling and discuss the potential of this system for producing squeezed states in the good cavity limit.

This paper is organised as follows. In section \ref{model}, we describe the model of the system and derive equations governing for the expectation values and variances. In section \ref{uncon}, we examine the unconditional dynamics and steady-state. In section \ref{cond}, we derive the conditional variances and examine the prospects for quantum squeezing. In section \ref{filter}, we provide the optimal filter for conditioning and in section \ref{rsb}, we examine the effect of resolved sideband cooling on the system.

\section{Model}\label{model}

\subsection{Parametric Drive}

Consider a mechanical mode of frequency $\omega_m$ in which the spring constant can be modulated at a frequency near $2\omega_m$. For dielectric oscillators, this is usually achieved using a high gradient electric field\cite{rugar,unterreithmeier}, but other methods exist\cite{suh,almog}. In position and momentum co-ordinates, the interaction Hamiltonian is
\begin{equation}
H = \frac{\hat p^2}{2m} + \frac{\hat x^2}{2}[k_0 + k_r \sin(\omega_d t + 2\theta)]\; ,
\end{equation}
where the parametric drive frequency $\omega_d = 2(\omega_m + \Delta)$.

\begin{figure}
\centering
\includegraphics[width=12cm]{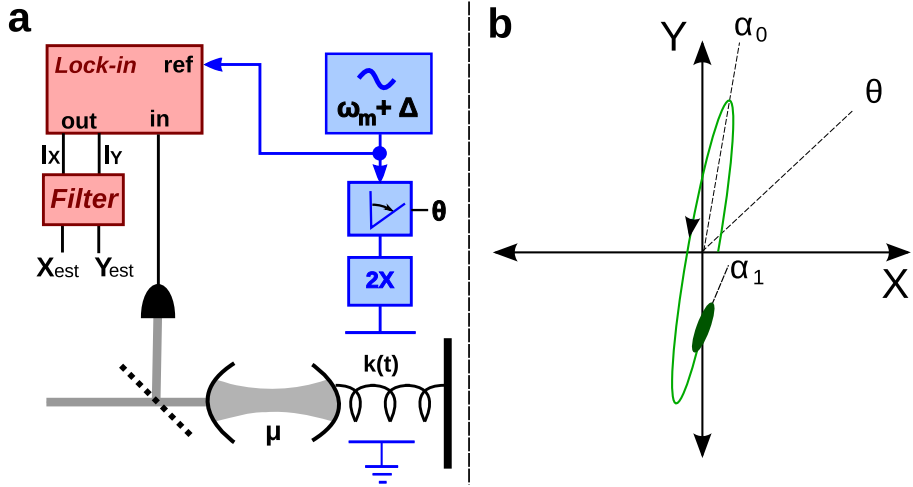}
\caption{\label{angles}a) Schematic of a optomechanical system with an electromechanical parametric drive. b) Illustration of an unconditional trajectory in X-Y phase space with a detuned drive applied near threshold, in the case $\Delta^2>\chi^2$. The shaded ellipse denotes the conditional uncertainty at a given time based on previous measurements. The angles $\theta=\pi/4$, $\alpha_0$ and $\alpha_1$ are defined between the Y axis and respective dotted lines.}
\end{figure}

For a high-Q oscillator, we can instead use as co-ordinates the quadratures $\hat X$ and $\hat Y$, the measurements of which are the outputs $I_X$ and $I_Y$ of a lock-in amplifier with a continuous position measurement as input (see Figure \ref{angles}a). These operators come from a rotating wave transformation at frequency $\omega_d/2$
\begin{equation}
\sqrt{\frac{m\omega_m}{\hbar}}\,\hat x = \hat X\sin(\omega_d t/2) + \hat Y\cos(\omega_d t/2)\; ,
\end{equation}
where, in terms of new creation and annihilation operators
\begin{eqnarray} 
\hat X &= (\hat a + \hat a^\dag)/\sqrt{2} \nonumber\\
\hat Y &= -i(\hat a - \hat a^\dag)/\sqrt{2} \nonumber \; ,
\end{eqnarray}
so that $[\hat X,\hat Y]=i$ and the ground-state variance for each quadrature is 1/2. The Hamiltonian can then be written in the rotating wave approximation as\cite{prl}
\begin{equation}\label{ham}
\tilde H = \hbar\Delta \hat a^\dag\hat a + i\hbar \frac{\chi}{2}(e^{2i\theta}\hat a^2-e^{-2i\theta}\hat a^{\dag2}) \; ,
\end{equation}
where $\chi=\omega_m k_r / 2k_0$, is approximately the peak-to-peak frequency modulation. Since we chose a rotating frame with respect to the parametric drive, this Hamiltonian looks like a resonance at $\Delta$ with a stationary squeezing operator. Note that the phase $\theta$ has no effect on the system dynamics but defines the squeezing axes with respect to the chosen quadratures.

\subsection{Measurement}
A stochastic master equation is used to model the measurement and damping\cite{jacobs}. Initially, we assume a cavity optomechanical system in the bad cavity limit ($\kappa \gg \omega_m$, where $\kappa$ is the cavity linewidth). We can define a measurement rate $\mu$ in this limit, given by
\begin{equation} \label{mu}
\mu=\frac{8g^2x_{zpf}^2\bar n}{\kappa} \; ,
\end{equation}
where $g$ is the optomechanical coupling rate, $\bar n$ is the mean photon number and $x_{zpf}=\sqrt{\hbar/ m\omega_m}$ is the RMS position due to zero-point motion. This parameter $\mu$ can be interpreted as a coupling rate to a zero-temperature measurement bath, which can be compared with the coupling rate to the thermal bath $\gamma$. The ratio $\mu/\gamma$ as well as the temperature will be used to define the various measurement regimes, as will be discussed in section \ref{cond}.

This kind of position measurement can be decomposed into quadratures. We will limit our analysis to the regime $\mu\ll\omega_m$, in which case the stochastic master equation can be similarly decomposed. The resulting master equation then resembles the well-studied model of heterodyne detection used in optical and microwave systems\cite{doherty}. Additionally, we assume the measurement signal has no thermal fluctuations. An observer's expected knowledge of the observable $A$ then evolves as
\begin{eqnarray}\label{master}
\fl \mathrm{d}\langle\hat A\rangle = -\frac{i}{\hbar}\langle[\hat A,\tilde H]\rangle\,\mathrm{d}t + [2\gamma N + \mu]\langle\mathcal{D}[\hat a^\dag]\hat A\rangle\,\mathrm{d}t + [2\gamma(N+1)+\mu]\langle\mathcal{D}[\hat a]\hat A\rangle\,\mathrm{d}t \\ 
 + \sqrt{\eta \mu}\langle\mathcal{H}[\hat X]\hat A\rangle\,\mathrm{d}W_1 + \sqrt{\eta\mu}\langle\mathcal{H}[\hat Y]\hat A\rangle\,\mathrm{d}W_2 \; . \nonumber
\end{eqnarray}
where $N$ is the mean bath phonon number, $\gamma=\omega_m/Q$ is the intrinsic damping rate, $\eta$ is the quantum efficiency and $\mathrm{d}W_1$ and $\mathrm{d}W_2$ are uncorrelated Wiener processes defining the residual noise given the measurement results $I_X=dQ_X/dt$ and $I_Y=dQ_Y/dt$
\begin{eqnarray} \label{dw1}
\mathrm{d}W_1 &=& \mathrm{d}Q_X - \sqrt{4\eta\mu}\langle\hat X\rangle\mathrm{d}t \\
\mathrm{d}W_2 &=& \mathrm{d}Q_Y - \sqrt{4\eta\mu}\langle\hat Y\rangle\mathrm{d}t \; . \label{dw2}
\end{eqnarray}
The superoperator
\begin{equation}
\mathcal{D}[\hat a]\hat A = \hat a^\dag\hat A\hat a - \frac{1}{2}(\hat a^\dag \hat a\hat A + \hat A\hat a^\dag \hat a)\; , \nonumber \\
\end{equation}
describes the thermal diffusion and back-action, while
\begin{equation}
\mathcal{H}[\hat a]\hat A = \hat a\hat A + \hat A\hat a^\dag - \langle \hat a+\hat a^\dag\rangle\langle\hat A\rangle \; . \nonumber
\end{equation}
describes the noise being introduced to the measurement record.

\subsection{Evolution of Observables}

Applying (\ref{master}) to the quadratures $X$ and $Y$ gives the evolution of the expectation values
\begin{eqnarray} \label{meanvalue1}
\fl\mathrm{d}\langle\hat X\rangle & = & [-(\gamma\!+\!\chi\cos 2\theta)\langle\hat X\rangle - (\Delta\!-\!\chi\sin 2\theta)\langle\hat Y\rangle]\mathrm{d}t + \sqrt{4\eta\mu}V_X\,\mathrm{d}W_1 + \sqrt{4\eta\mu}C\,\mathrm{d}W_2 \\
\fl\mathrm{d}\langle\hat Y\rangle & = & [-(\gamma\!-\!\chi\cos 2\theta)\langle\hat Y\rangle + (\Delta\!+\!\chi\sin 2\theta)\langle\hat X\rangle]\mathrm{d}t + \sqrt{4\eta\mu}C\,\mathrm{d}W_1 + \sqrt{4\eta\mu}V_Y\,\mathrm{d}W_2 \label{meanvalue2}
\end{eqnarray}
where $V_X$ and $V_Y$ are the variances in the $X$ and $Y$ quadratures respectively, and we define the covariance $C=\langle \hat X \hat Y + \hat Y\hat X \rangle/2 - \langle\hat X\rangle\langle\hat Y\rangle$. Setting $\Delta=\mu=\theta=0$, this describes an additional damping in $X$ and reduced damping in $Y$, both proportional to the drive strength $\chi$. In general, the system is stable if $\chi$ is below a threshold value
\begin{equation} \label{thresh}
\chi_{th} = \sqrt{\Delta^2 + \gamma^2} \; .
\end{equation}
The detuning causes a continuous rotation between amplified and damped quadratures, thus enabling a stronger parametric drive without self-oscillation. For $\chi^2 < \Delta^2$, the oscillator's trajectories through X-Y phase space are elliptical with frequency
\begin{equation}
\omega_e = \sqrt{\Delta^2 - \chi^2}\; ,
\end{equation}
and for $\chi_{th}^2 > \chi^2 > \Delta^2$, the trajectories take the form $[c_1\sinh(i\omega_e) + c_2\cosh(i\omega_e)]e^{-\gamma t}$.

We can also find the evolution of the variances by applying the master equation to $\hat X^2$, $\hat Y^2$ and $\hat X\hat Y+\hat Y\hat X$, then using
\begin{eqnarray}
\mathrm{d}V_A &=& \mathrm{d}\langle \hat A^2\rangle - 2\langle \hat A\rangle\mathrm{d}\langle \hat A\rangle - (\mathrm{d}\langle \hat A\rangle)^2 \nonumber \\
\mathrm{d}C &=& \frac{1}{2}\mathrm{d}\langle \hat X\hat Y\!+\!\hat Y\hat X\rangle \!-\! \langle \hat X\rangle\mathrm{d}\langle \hat Y\rangle \!-\! \langle \hat Y\rangle\mathrm{d}\langle \hat X\rangle \!-\! \mathrm{d}\langle \hat X\rangle\mathrm{d}\langle \hat Y\rangle \; . \nonumber
\end{eqnarray}
Applying the It\=o rules (i.e.\ $(dt)^2=dWdt=0$ and $(dW)^2=dt$) and setting third and higher order cumulants to zero (i.e.\ Gaussian input states) produces quadratic differential equations for the variances
\begin{eqnarray} \label{arbphase1}
\fl \frac{d}{dt}V_X & = & -2(\gamma\!+\!\chi\cos(2\theta)) V_X - 2(\Delta\!-\!\chi\sin(2\theta))C + \gamma(2N\!+\!1)\!+\!\mu - 4\eta\mu(V_X^2+C^2)  \\
\fl \frac{d}{dt}V_Y & = & -2(\gamma\!-\!\chi\cos(2\theta)) V_Y + 2(\Delta\!+\!\chi\sin(2\theta))C + \gamma(2N\!+\!1)\!+\!\mu - 4\eta\mu(V_Y^2+C^2) \label{arbphase2} \\
\fl \frac{d}{dt}C & = & -2\gamma C - \Delta(V_Y\!-\!V_X) + \chi\sin(2\theta)(V_X\!+\!V_Y) - 4\eta\mu C(V_X\!+\!V_Y) \; . \label{arbphase3}
\end{eqnarray}

\section{Unconditional Steady-State}\label{uncon}

Let us for now take the limit where the efficiency $\eta=0$, implying that all of the measurement results are discarded. This gives us the unconditional dynamics, and the variances obtained are those that would be inferred from a spectrum analysis over an infinite time. In this case, the measurement strength only appears as an additional phonon number due to back-action
\begin{equation}
N_{BA} = \frac{\mu}{2\gamma} \; .
\end{equation}
Applying detuning shifts the angle of the squeezing axes significantly. In anticipation of this we set $\theta = \pi/4$, which without detuning will amplify fluctuations maximally along an axis rotated through an angle $\pi/4$ from the $Y$ quadrature (Note: this will henceforth be called the \emph{antisqueezing angle}). Solving (\ref{arbphase1}-\ref{arbphase3}) with the time derivatives equal to zero results in the steady-state variances
\begin{eqnarray} \label{unconditional1}
V_{X} & = & \left(1-\frac{\chi(\Delta-\chi)}{\gamma^2+\Delta^2-\chi^2}\right)\, (V_T+N_{BA})\\
V_{Y} & = & \left(1+\frac{\chi(\Delta+\chi)}{\gamma^2+\Delta^2-\chi^2}\right)\, (V_T+N_{BA})\\ \label{unconditional2}
C & = & \frac{\chi \gamma}{\gamma^2+\Delta^2-\chi^2} \, (V_T+N_{BA}) \; ,
\end{eqnarray}
where $V_T = N + 1/2$. Clearly, for $\Delta\approx\chi$, the antisqueezing axis is close to $Y$ as it would be for $\theta=\Delta=0$.

The exact squeezing is characterized by minimizing the variance over an angle $\alpha$ where
\begin{equation}
V_{\alpha} = V_X \cos^2 \alpha + V_Y\sin^2 \alpha - 2C \cos\alpha\sin\alpha \; .
\end{equation}
As long as $V_Y>V_X$, the maximally squeezed and antisqueezed quadratures ($V_-$ and $V_+$) become
\begin{eqnarray}
V_\pm &=& \frac{1}{2} [(V_X+V_Y) \pm (V_Y-V_X)\sec(2\alpha_0)] \qquad \mathrm{where} \\
\alpha_0 &=& \frac{1}{2} \tan^{-1} \left(\frac{2C(t)}{V_Y(t)-V_X(t)}\right)\; ,
\end{eqnarray}
where $\alpha_0$ is the antisqueezing angle (see Figure \ref{angles}b). Applying this to Eqs (\ref{unconditional1}-\ref{unconditional2}) gives
\begin{eqnarray} \label{general1}
V_{-} &=& \frac{V_T+N_{BA}}{1+\chi/\chi_{th}}\\
V_{+} &=& \frac{V_T+N_{BA}}{1-\chi/\chi_{th}} \label{general2}\\
\alpha_{0} &=& \frac{1}{2}\tan^{-1}(\frac{\gamma}{\Delta}) \label{general3} \; ,
\end{eqnarray}
where the threshold value $\chi_{th}$, given by (\ref{thresh}) defines the maximum drive strength before the system self-oscillates. Therefore, for both detuned and resonant drives, a maximum unconditional squeezing $V_-/V_T$ of $-3$dB can be achieved. Squeezing that surpasses this amount has been reported in the literature\cite{rugar,suh}, however this was measured using only frequency components of the motion near the mechanical resonance. When only the resonance is included, the maximum squeezing can be confirmed by using Langevin equations in the Fourier domain to be 6dB\cite{collett}. This is not in contradiction with our time-domain approach, which is effectively an integral over all frequency components.

Note that the squeezing we have discussed in this section is relative to the thermal variance, $V_-/V_T$, which will for our purposes be called \emph{classical squeezing}. To achieve \emph{quantum squeezing}, the variance must be below the Heisenberg limit. The quantum squeezing ratio is then $V_-/V_g$, where $V_g$ in our units equals 1/2. For quantum squeezing of the unconditional variance, the following condition is required
\begin{equation}
N + N_{BA} < 1/2 \; ,
\end{equation}
which places a strict upper bound on the measurement strength and temperature.

\section{Conditional Steady-State}\label{cond}

We will now demonstrate that if $\eta>0$, using a detuned parametric drive enables a much greater degree of classical and quantum squeezing. This is because the squeezed quadrature can be inferred from previous measurements of the amplified quadrature. The conditional variance arising from the master equation can be thought of as the mean-square difference between the quadrature amplitude $X$ the observer's optimal estimate $X_{est}$
\begin{equation}
V_X = \langle (\hat X(t) - X_{est}(t))^2 \rangle \; ,
\end{equation}
and similarly for $Y$. This assuming that $X_{est}$ and $Y_{est}$ are calculated in the correct way from the measurement record. We will derive this optimal filter in the next section.

With no parametric drive, the steady-state conditional variance in both quadratures can be obtained by simply setting $dV/dt=0$ in (\ref{arbphase1}-\ref{arbphase3})
\begin{equation}\label{meas}
V_0 = \frac{\sqrt{1 + 4\mathrm{SNR}} - 1}{4\eta\mu/\gamma}\; ,
\end{equation}
where
\begin{equation}
\mathrm{SNR} = 2\eta\mu(V_T + N_{BA})/\gamma = 2\eta\mu(N+1/2)/\gamma + \frac{\eta\mu^2}{\gamma^2} \; ,
\end{equation}
is a ratio of mechanical signal to shot-noise. In the classical regime of large $N$, this can be thought of as the ratio of mean-square thermal displacement $(\delta x_T)^2$ to $(\delta x_\gamma)^2$ the square of the minimum distance resolvable over a time $\Delta t\approx 1/(4\gamma)$\cite{doherty}

Taking $\mathrm{SNR} \ll 1$ in Eq.\ (\ref{meas}) produces the bad measurement regime, at the limit of which $V_0$ reduces to the unconditional variance $V_T$. As SNR increases past unity so that $1/(2\eta V_T) < \mu/\gamma < 2V_T$, the conditional variance reduces towards the ground state. When $\mu/\gamma \gg 2V_T$, the back-action in SNR (proportional to $(\mu/\gamma)^2$) dominates and the strong measurement limit is approached ($V_0\rightarrow1/(2\sqrt{\eta}))$. Note that when $2V_T=\eta=1$, the second regime (i.e.\ classical measurement) disappears entirely.

When performing position estimation, a detuned parametric drive results in an elliptical gaussian uncertainty. However, the antisqueezing angle of this distribution does not in general correspond to $\alpha_0$, which defines the axis of the average elliptical trajectory. Figure \ref{angles}b illustrates this difference, where a parametric drive phase of $\pi/4$ results in an antisqueezing angle $\alpha_1$ for the conditional variance.

In order to solve the variances for the maximally squeezed and antisqueezed quadratures, it is convenient to define the pump phase $\theta$ as a function of SNR and other pump parameters so that the $X$ quadrature is always maximally squeezed and the covariance vanishes. In order to work in terms of the equivalent antisqueezing angle, we make the replacement $\theta = \pi/4 - \alpha_1$. Applying the steady-state condition and $C=0$ to equations (\ref{arbphase1}-\ref{arbphase3}) then produces
\begin{eqnarray}\label{cond1}
V_X &=& \frac{\sqrt{(\gamma+\chi\sin(2\alpha_1))^2+4\gamma^2 \mathrm{SNR}}-\gamma-\chi \sin(2\alpha_1)}{4\eta\mu}\\
V_Y &=& \frac{\sqrt{(\gamma-\chi\sin(2\alpha_1))^2+4\gamma^2 \mathrm{SNR}}-\gamma+\chi \sin(2\alpha_1)}{4\eta\mu} \label{cond2} \\
\cos (2\alpha_1) &=& \frac{\Delta(V_Y-V_X)}{\chi(V_Y+V_X)} \label{cosalpha} \; .
\end{eqnarray}

\begin{figure}[!tb]
\centering
\includegraphics[width=10cm]{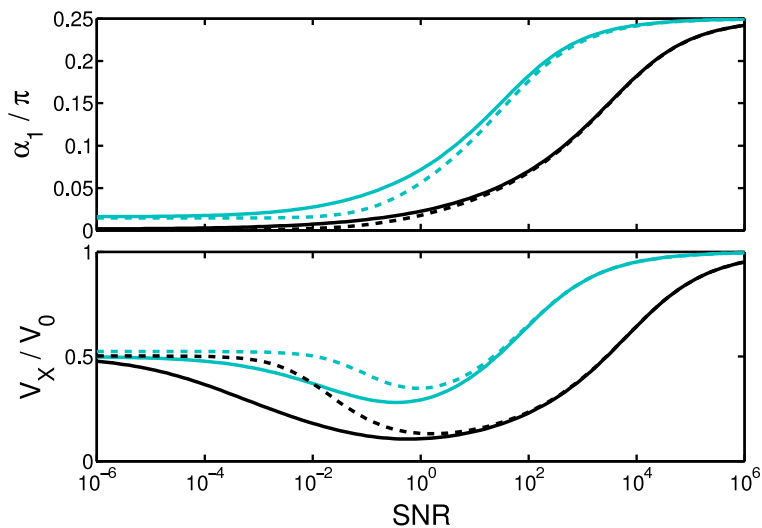}
\caption{\label{snr}(Above) Antisqueezing angle $\alpha_1$ and (below) squeezing ratio $V_X/V_0$ vs SNR. Light curves are for a drive strength $\chi=10\gamma$ while dark curves are for $\chi=100\gamma$. Solid lines are for the pump detuning on threshold, while dashed lines indicate detuning away from threshold by $\gamma$. Note that the squeezing disappears altogether in the strong measurement limit.}
\end{figure}

The conditional variances may be solved exactly by finding the antisqueezing angle $\alpha_1$ in terms of system parameters. The antisqueezing angle (derived in \ref{appen}) satisfies
\begin{equation} \label{alphasol}
\fl \cos2\alpha_1\!=\! \frac{\Delta}{\chi_{th}}\!\left(\!\frac{\chi_{th}^2\!+\!\chi^2\!+\!4\gamma^2\mathrm{SNR}\!-\!\sqrt{(\chi_{th}^2\!-\!\chi^2)^2\!+\!8(\chi_{th}^2\!+\!\chi^2)\gamma^2\mathrm{SNR}\!+\!16\gamma^4\mathrm{SNR}^2}}{2\chi^2}\right)^{\frac{1}{2}}
\end{equation}

\begin{figure}[!tb]
\centering
\includegraphics[width=10cm]{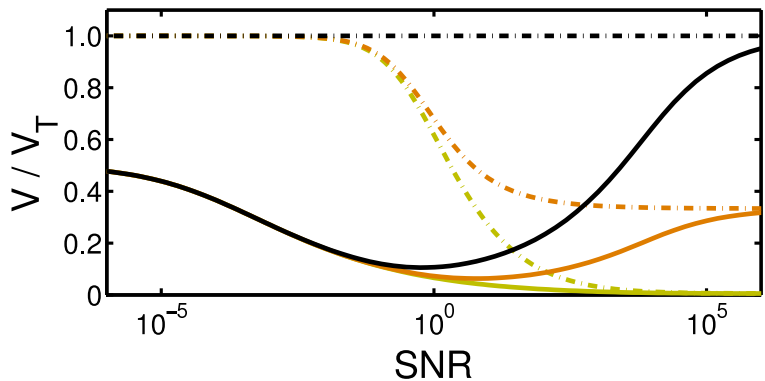}
\caption{\label{snr2}Conditional variance $V$ normalised to thermal variance $V_T$, for a quadrature with no parametric drive ($V_0$, dotted lines) and in the squeezed quadrature with the drive $\chi=100\gamma$ and detuning at threshold ($V_X$, solid lines). Initial phonon numbers are $N=100$ (light), $N=1$ (medium) and $N=0$ (dark). The increase in squeezed variance at high SNR and low temperature demonstrates the effect of back-action on quantum squeezing.}
\end{figure}

From examining the above, we see that the antisqueezing angle increases from the unconditional angle $\alpha_0$ at $\mathrm{SNR}=0$ up to $\pi/4$ in the limit $\mathrm{SNR} \rightarrow \infty$. Inserting this into (\ref{cond1}-\ref{cond2}) gives the squeezed and antisqueezed steady-state variances. Notably, since $\alpha_1$ does not depend on $N$, $\eta$ and $\mu$ separately but on their combined form SNR, the ratio of squeezed conditional variance to the undriven conditional variance $V_X/V_0$ has the same property. These dependences on SNR are plotted in Figure \ref{snr}. This \emph{conditional squeezing} ratio starts at the unconditional squeezing ($\approx1/2$), drops to a minimum near $\mathrm{SNR}\approx1$, and approaches 1 in the strong measurement limit.

The degradation of the conditional squeezing for strong measurement occurs even in the classical regime where backaction heating is negligible. Once the measurement sufficiently strong to directly resolve the squeezed quadrature, the additional benefit provided by parametric squeezing is reduced. This can be seen by the squeezed and antisqueezed variances (\ref{cond1}-\ref{cond2}) becoming independent of the parametric drive as $\mathrm{SNR}\rightarrow\infty$. The effect of the parametric drive in the squeezed quadrature can therefore be interpreted as a signal boost from the amplified quadrature, which is of greatest benefit near $\mathrm{SNR}\approx1$ where the conditional variance $V_0$ is most sensitive to SNR. This is supported by the fact that the squeezed conditional variance $V_X$ starts to significantly reduce at a lower SNR than the bare conditional variance $V_0$, as illustrated in Figure \ref{snr2}.

At threshold, Eq.\ (\ref{alphasol}) reduces to
\begin{equation}
\cos(2\alpha_1) = \frac{\sqrt{\chi^2-\gamma^2}}{\chi^2}\, \left(\sqrt{\chi^2+\gamma^2\mathrm{SNR}}-\gamma\sqrt{\mathrm{SNR}}\right) \; .
\end{equation}
In the limit $\chi \gg \gamma$ and $\mathrm{SNR}=1$, the following approximation can then be made
\begin{equation}
\chi\sin(2\alpha_1) \approx \gamma\sqrt{2\chi/\gamma} \; .
\end{equation}
The amount of squeezing (quantum or classical) achievable is similarly proportional to $\sqrt{\chi/\gamma}$, which is limited only by the rotating wave approximation to be less than the square root of the oscillator Q factor (i.e.\ $\chi \ll \omega_m$).

\begin{figure}[t]
\centering
\includegraphics[width=10cm]{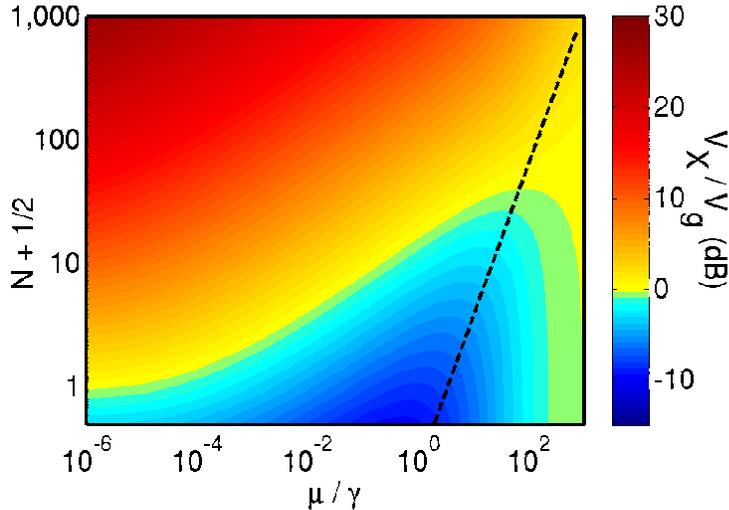}
\caption{\label{quant}Colour plot of the ratio of squeezed variance to ground-state variance for $\chi=\Delta=100\gamma$. A ratio less than unity indicates quantum squeezing. The dashed line indicates $\mu/\gamma = N+1/2$, beyond which back-action is dominant.}
\end{figure}

At an initial ground state, the optimal regime $\mathrm{SNR}\approx1$ is only on the cusp of the back-action dominated regime (assuming the efficiency is near unity). To illustrate the effect of the parametric drive and estimation at low temperature, $V_X$ is plotted for various $\mu$ and low values of $N$ in Figure \ref{quant}. Squeezing of the conditional variance below the zero-point motion is achieved when $V_X<V_g=0.5$, which is possible even from relatively high initial temperatures or with inefficient detection. The appearance of an optimum measurement strength $\mu$ at low temperature is in stark contrast with the best possible squeezing using a resonant drive, which degrades steadily from a minimum $0.5V_g$ at $\mu=0$ to $0.73V_g$ at $\mu=\gamma$. Even more notably, this scheme vastly outperforms back-action evasion in this parameter regime\cite{prl}.

\section{Optimal Filter}\label{filter}

The idea of optimal estimation is based on the fact that since the mechanical noise is weighted by a Lorentzian susceptibility, it can be partially filtered out from the uniform shot-noise. In real-time control situations such as feedback cooling, the stochastic master equation (\ref{master}) can be used to derive a general time-dependent Kalman filter for the position estimate\cite{doherty2}. We are interested in a simple filter which produces the minimum conditional variances as obtained in the previous section. This can be achieved by applying the steady-state condition to the master equation. In this case the measurement signals as defined in Eqs (\ref{dw1}-\ref{dw2}) are related to the estimate as follows\cite{milburn}
\begin{eqnarray}
\mathrm{d}Q_X &=& \sqrt{4\eta\mu}\langle X_{est}\rangle \mathrm{d}t + \mathrm{d}W_1 \\
\mathrm{d}Q_Y &=& \sqrt{4\eta\mu}\langle Y_{est}\rangle \mathrm{d}t + \mathrm{d}W_2 \; .
\end{eqnarray}
Using the above, we can rewrite the mean value equations Eqs (\ref{meanvalue1}-\ref{meanvalue2}) in terms of the measurement rather than the white noise residual $dW$. This results in additional damping terms proportional to the measurement rate $\eta\mu$. As in the previous section, we will choose a pump phase $\theta=\pi/4-\alpha_1$ and set $C=0$. Substituting these values, Fourier transforming and solving for $\langle X_{est}(\omega)\rangle$ gives
\begin{equation} \label{filterfreq}
\fl \langle X_{est}(\omega)\rangle = \sqrt{4\eta\mu}\frac{(\gamma_Y - \chi\sin(2\alpha_1) + i\omega)V_XQ_X(\omega)-(\Delta-\chi\cos(2\alpha_0))V_YQ_Y(\omega)}{\Delta^2-\chi^2+\gamma_X\gamma_Y +\chi\sin(2\alpha_1)(\gamma_Y-\gamma_X) - \omega^2 + i\omega(\gamma_X+\gamma_Y)}
\end{equation}
where the intrinsic and measurement-induced damping rates of the $X$ and $Y$ quadratures are respectively combined as
\begin{eqnarray}
\gamma_X &=& \gamma + 4\eta\mu V_X \\ 
\gamma_Y &=& \gamma + 4\eta\mu V_Y \; .
\end{eqnarray}
This specifies an increasing filter bandwidth as the mechanical signal overtakes the shot-noise. These quantities are equal in the bad measurement limit (where $\eta\mu\rightarrow 0$) as well as the strong measurement limit (where $V_X \approx V_Y \approx 1/(2\sqrt{\eta})$), but vastly different near $\mathrm{SNR}\approx 1$ if driven near threshold. We will limit ourselves to the regime $\Delta^2>\chi^2$, where the general form of the above solution in the time domain is
\begin{equation}
\fl \langle X_{est}(t)\rangle = g_{XX}Q_X(t) * \left[\cos(\Omega t - \phi)e^{-\Gamma t}\right] + g_{XY}Q_Y(t) * \left[\sin(\Omega t)e^{-\Gamma t}\right] \; ,
\end{equation}
and similarly for the $Y$ estimate
\begin{equation}
\fl \langle Y_{est}(t)\rangle = g_{YY}Q_Y(t) * \left[\cos(\Omega t + \phi)e^{-\Gamma t}\right] + g_{YX}Q_X(t) * \left[\sin(\Omega t)e^{-\Gamma t}\right] \; .
\end{equation}
The parameters are obtained from (\ref{filterfreq}) and simplified in terms of $\alpha_1$ using relations from \ref{appen}.
\begin{eqnarray}
\Omega &=& \sqrt{\Delta^2-\chi^2\cos^2(2\alpha_1)(1+\gamma^2/\Delta^2)} \\
\Gamma &=& \frac{1}{2}(\gamma_X + \gamma_Y) \nonumber \\
 &=& \Delta\tan(2\alpha_1) \\
\phi &=& \tan^{-1} \left(\frac{\chi\gamma\cos(2\alpha_1)}{\Delta\Omega}\right) \\
g_{XX} &=& \sec\phi\sqrt{4\eta\mu}V_X \\
g_{XY} &=& \frac{\Delta-\chi \cos(2\alpha_1)}{\Omega}\sqrt{4\eta\mu}V_Y \\
g_{YY} &=& \sec\phi\sqrt{4\eta\mu}V_Y \\
g_{YX} &=& \frac{\Delta+\chi \cos(2\alpha_1)}{\Omega}\sqrt{4\eta\mu}V_X \; .
\end{eqnarray}
These expressions simplify in the $\mathrm{SNR}=0$ limit (where $\cos(2\alpha_1)=\Delta/\chi_{th}$) to expected values (e.g.\ $\Omega=\omega_e$ and $\Gamma = \gamma$). In the high SNR limit (where $\cos(2\alpha_1) = 0$) the optimal filter has infinite bandwidth and is independent of the parametric drive.

The measurement records for X and Y, upon applying these Lorentzian filters, form an optimal estimate of the oscillator's current position in phase space. In other words, the amplitudes of the in-phase and out-of-phase fluctuations are known to within uncertainties defined by $V_X$ and $V_Y$. By mixing the estimate back up to $\omega_d/2$ and with the correct phase shift, an appropriate feedback cooling signal is obtained. In this way, conditional squeezing is turned into real squeezing\cite{wiseman}.

\section{Resolved Sideband Cooling}\label{rsb}

Up to this point we have only considered optomechanical systems as ultra-sensitive transducers of position fluctuations. It is natural to ask whether there is any advantage in using a detuned parametric drive in conjunction with the near-ubiquitous optomechanical technique of sideband cooling. The resonant driving case of this has been analysed previously, albeit with a focus on the cavity output spectrum\cite{woolley}.

We can model sideband cooling by extending the master equation to include a cavity bath. This is done by adiabatically eliminating the cavity\cite{gardiner}, and results in the back-action terms in equation (\ref{master}) being replaced by terms analogous to those for the thermal bath. The deterministic part of the master equation is then
\begin{eqnarray}
\fl \langle\mathrm{d}\langle\hat A\rangle\rangle &=& -\frac{i}{\hbar}\langle[\hat A,\tilde H]\rangle\,\mathrm{d}t + 2\gamma N\langle\mathcal{D}[\hat a^\dag]\hat A\rangle\,\mathrm{d}t + 2\gamma(N+1)\langle\mathcal{D}[\hat a]\hat A\rangle\,\mathrm{d}t \\ 
\fl & & + 2\gamma_C N_C\langle\mathcal{D}[\hat a^\dag]\hat A\rangle\,\mathrm{d}t + 2\gamma_C(N_C+1)\langle\mathcal{D}[\hat a]\hat A\rangle\,\mathrm{d}t \; , \nonumber
\end{eqnarray}
where the general forms of the optical damping $\gamma_C$ and effective cavity temperature $N_C$ are given in \cite{marquardt} as $\Gamma_{opt}$ and $\bar n^O_M$. Taking the limit of large cavity loss $\kappa \gg \omega_m$ and zero cavity detuning, the back-action noise $\gamma_C(2N_C+1)$ from the master equation is equal to $\mu$ as we have defined in Eq.\ (\ref{mu}) so our approach is consistent with \cite{marquardt}.

On the red sideband (i.e.\ cavity detuning equal to $-\omega_m$) and in the good cavity limit, the cavity temperature $N_C$ approaches zero and all photons detected are a product of phonon absorption. Consequently, only downgoing transitions appear in the measurement terms of the master equation. In this regime, adiabatic elimination can be performed on the stochastic master equation for the coupled cavity-oscillator system as in \cite{doherty2}. The resulting stochastic master equation for the oscillator alone is then equivalent to heterodyne detection of a cavity output\cite{milburn} and has the form
\begin{eqnarray}
\fl \mathrm{d}\langle\hat A\rangle &=& -\frac{i}{\hbar}\langle[\hat A,\tilde H]\rangle\,\mathrm{d}t + [2\gamma N]\langle\mathcal{D}[\hat a^\dag]\hat A\rangle\,\mathrm{d}t + [2\gamma(N+1) + \mu]\langle\mathcal{D}[\hat a]\hat A\rangle\,\mathrm{d}t \\ 
\fl & & + \sqrt{\eta\mu/2}\langle\mathcal{H}[\hat a]\hat A\rangle\,\mathrm{d}W_1 + \sqrt{\eta\mu/2}\langle\mathcal{H}[i\hat a]\hat A\rangle\,\mathrm{d}W_2 \; . \nonumber
\end{eqnarray}
In this case, the Wiener processes are
\begin{eqnarray}
\mathrm{d}W_1 &=& \mathrm{d}Q_X - \sqrt{\eta\mu}\langle\hat X\rangle\mathrm{d}t \\
\mathrm{d}W_2 &=& \mathrm{d}Q_Y - \sqrt{\eta\mu}\langle\hat Y\rangle\mathrm{d}t \; .
\end{eqnarray}
This master equation leads to variance equations by the same method as section \ref{model}. Letting $\theta=\pi/4$
\begin{eqnarray}
\fl\frac{d}{dt}V_X & = & -(2\gamma+\mu)(V_X-1/2) - 2(\Delta-\chi)C + 2N\gamma - \eta\mu[(V_X-1/2)^2+C^2] \\
\fl\frac{d}{dt}V_Y & = & -(2\gamma+\mu)(V_Y-1/2) + 2(\Delta+\chi)C + 2N\gamma - \eta\mu[(V_Y-1/2)^2+C^2] \\
\fl\frac{d}{dt}C & = & -(2\gamma+\mu) C - \Delta(V_Y-V_X) + \chi(V_X+V_Y) - \eta\mu C(V_X+V_Y-1) \; .
\end{eqnarray}
Compared to the standard continuous measurement derived earlier, the additional terms proportional to $\mu$ here are an unconditional linear damping as well as an offset in the conditioning term, such that all measurement terms disappear for a symmetric pure state $V_X=V_Y=1/2$ with $C=0$.
The threshold condition is now dependent on $\mu$
\begin{equation}
\chi_{th}^{RSB} = \sqrt{(\gamma + \mu/2)^2 + \Delta^2} \; ,
\end{equation}
and the unconditional variance with no parametric drive is
\begin{equation}
V_T = \frac{2\gamma}{2\gamma+\mu}N + \frac{1}{2} \; ,
\end{equation}
as in \cite{marquardt}. For perfect efficiency, the linear optical damping terms vanish since
\begin{equation}
-\mu(V_X-1/2) - \mu(V_X-1/2)^2 = -\mu V_X^2 + \mu/4 \; ,
\end{equation}
leaving variance equations that are identical to (\ref{arbphase1}-\ref{arbphase3}) apart from an expected factor of 4 in measurement strength\cite{schliesser}. However, for $\eta \ll 1$, resolved sideband cooling offers a qualitative difference since a pure state is always approached in the strong measurement limit. Without cooling, this limiting variance increases by a factor of $1/\sqrt{\eta}$.

\begin{figure}[bt]
\centering
\includegraphics[width=14cm]{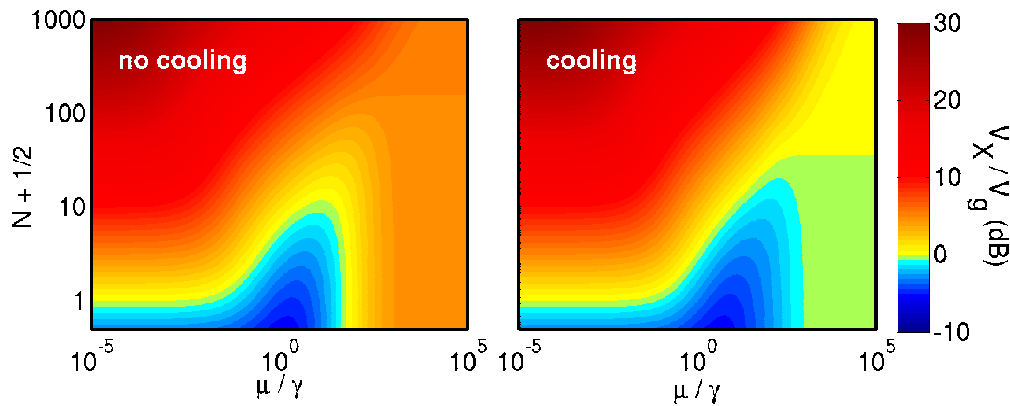}
\caption{\label{rsbfig}Squeezed conditional variance for an inefficient detector ($\eta=0.1$), normalised to the ground state variance, in the (left) normal and (right) ideal resolved sideband regime. In both cases, $\chi=100\gamma$ and detuning from threshold is $\gamma$.}
\end{figure}

These differences are small, however, in the important regime around $\mu\approx\gamma$. This is shown in Figure \ref{rsbfig}, using numerical solutions to the steady-state variance equations. With resolved sideband cooling, the squeezing extends further into the back-action dominated region $\mu>\gamma$, however the maximum squeezing is unchanged. Outside the good cavity limit, the cavity bath temperature increases and the result can be expected to be worse. Therefore, while resolved sideband cooling aids in resolving zero-point fluctuations in the presence of detector inefficiency, the requirement of strong measurement precludes it from being significantly useful in the context of parametric squeezing. Sideband cooling is, however, quite compatible with backaction evasion-based squeezing schemes as they both operate best in the (would-be) back-action dominated regime and can both be implemented using similar techniques\cite{hertzberg}.

\section{Conclusion}

Unlike classical systems, the strength of any continuous measurement of a quantum system plays an important role in its dynamics, with the optimal strength depending on the application. Many uses are being found for weak measurement\cite{simon}, while quantum non-demolition techniques such as back-action evasion work best in the strong measurement limit\cite{clerk,gangat}. We have described one case in which a measurement strength near the standard quantum limit is preferred, as it is in gravity-wave detectors for similar reasons\cite{caves}.

We have shown that with optimal estimation, a detuned parametric drive greatly reduces the uncertainty of one quadrature of motion when the signal-to-noise ratio is of order unity. At low temperature, since the back-action in this regime is weak, the parametric drive allows uncertainties well below the ground state. To our knowledge, this is the only steady-state scheme for parametric squeezing for which the squeezing is limited solely by device parameters and not by fundamental constraints.

We have derived the optimal filter for producing the position estimate, which can be used to perform feedback and achieve single quadrature confinement to below ground-state uncertainty. Confinement based on resolved-sideband cooling only has a significant effect in the back-action resolved regime, which is not conducive to squeezing by parametric methods.

\ack
This research was funded by the Australian Research Council Centre of Excellence CE110001013 and Discovery Project DP0987146.

\clearpage
\appendix

\section{Derivation of Conditional Squeezing Angle}
\label{appen}
With the steady-state and $C=0$ conditions, the quadratic equations (\ref{arbphase1}-\ref{arbphase2}) can be expressed as
\begin{equation} \label{equality}
\fl 2\eta\mu V_Y^2 + (\gamma-\chi\sin(2\alpha_1))V_Y = 2\eta\mu V_X^2 + (\gamma+\chi\sin(2\alpha_1))V_X = \gamma(V_T+N_{BA}) \; .
\end{equation}
Rearranging the first equality,
\begin{equation} \label{equality2}
2\eta\mu(V_Y+V_X)(V_Y-V_X) + \gamma(V_Y-V_X) = \chi \sin(2\alpha_1)(V_Y+V_X) \; .
\end{equation}
Dividing this through by $V_Y-V_X$ and using Eq.\ (\ref{cosalpha}) gives a new equation for the antisqueezing angle
\begin{equation} \label{tanalpha}
\Delta\tan(2\alpha_1) = 2\eta\mu(V_Y + V_Y) + \gamma \; ,
\end{equation}
which as expected, reduces to the unconditional result Eq.\ (\ref{general3}) when $\eta\mu=0$. We would like a form for $\alpha_1$ in terms of system parameters only, for which we can rearrange Eq.\ (\ref{equality}) again in the form
\begin{equation}
\fl 2\gamma(V_T + N_{BA}) = 2\eta\mu (V_Y^2+V_X^2) + \gamma(V_Y+V_X) - \chi\sin(2\alpha_1)(V_Y-V_X) \; .
\end{equation}
Dividing (\ref{equality2}) through by $V_Y+V_X$, then substituting for $\chi\sin(2\alpha_1)$ above leaves a drive-independent relation between the two quadrature variances
\begin{equation}\label{quadrelation}
2V_YV_X\left(\eta\mu+\frac{\gamma}{V_Y+V_X}\right) = \gamma (V_T+N_{BA}) \; .
\end{equation}
Substituting the relation (from Eq.\ (\ref{cosalpha}))
\begin{equation}
1-\frac{\chi^2}{\Delta^2}\cos^2(2\alpha_1) = \frac{4V_Y V_X}{(V_Y+V_X)^2} \; ,
\end{equation}
as well as Eq.\ (\ref{tanalpha}) into (\ref{quadrelation}), we can obtain an expression containing only a function of $\alpha_1$, the thermal variance $V_T$ and other parameters. The general form of $\alpha_1$ is now the solution to the equation
\begin{equation}
\left(\Delta^2\tan^2(2\alpha_1)-\gamma^2\right)\left(1-\frac{\chi^2}{\Delta^2}\cos^2(2\alpha_1)\right) = 8\eta\mu\gamma(V_T+N_{BA}) \; ,
\end{equation}
or written in terms of SNR,
\begin{equation}
\Delta^2\tan^2(2\alpha_1) - \chi^2\sin^2(2\alpha_1) + \frac{\chi^2\gamma^2}{\Delta^2}\cos^2(2\alpha_1) = \gamma^2(1+4\mathrm{SNR}) \; .
\end{equation}
This can be solved analytically by multiplying through by $\cos^2(2\alpha_1)$ and transforming sine to cosine, then solving a quadratic equation, resulting in
\begin{equation}
\fl \cos2\alpha_1\!=\! \frac{\Delta}{\chi_{th}}\!\left(\!\frac{\chi_{th}^2\!+\!\chi^2\!+\!4\gamma^2\mathrm{SNR}\!-\!\sqrt{(\chi_{th}^2\!-\!\chi^2)^2\!+\!8(\chi_{th}^2\!+\!\chi^2)\gamma^2\mathrm{SNR}\!+\!16\gamma^4\mathrm{SNR}^2}}{2\chi^2}\right)^{\frac{1}{2}}
\end{equation}

\section*{References}

\providecommand{\newblock}{}

\end{document}